\documentclass[a4,amsmath,amssymb]{revtex4}
\usepackage{amsmath}
\usepackage{amssymb}
\usepackage{graphicx}% Include figure files
\usepackage{bm}% bold math

\def\eqnn#1{(\ref{eq:#1})}

\def\figno#1{Fig.~\ref{fig:#1}}

\def\Im{{\rm \,Im}}

\def\vev#1{\langle #1\rangle}
\def\Im{{\rm \,Im}}

\def\kb{k_{\scriptscriptstyle\rm B}}

\def\bsize{b}

\def\spec{S_{\rm h}}
\begin{document}
\title{
  Michelson interferometry with quantum noise reduction
}
\author{Takahisa
  Mitsui 
 and Kenichiro Aoki
 }
\affiliation{Research and Education Center for Natural Sciences and
  Dept. of Physics, Hiyoshi, Keio University, Yokohama 223--8521,
  Japan}
\begin{abstract}
    A Michelson interferometer with noise reduction to sub-shot noise
    levels is proposed and realized. Multiple measurements of a single
    signal beam are taken and the quantum property of light plays an
    essential role in the principle underlying this interferometry.
    The method makes use of the coherent state of light and requires
    only a simple modification to the standard Michelson
    interferometer.  The surface fluctuation spectra of liquids are
    measured using this method down to a few orders of magnitude below
    the shot noise level.  The spectrum derived from hydrodynamical
    considerations agrees well with the observed results for
    water. However, for oil, slight deviations are seen at high
    frequencies ($\gtrsim1\,$MHz), perhaps indicating its more complex
    underlying physics.  The measurement requires a relatively low
    light power and a short time, so that it has a wide range of
    applicability.
\end{abstract}
% \vspace{3mm}
\maketitle
% While the quantum nature of light has been established and studied for
% some time, there have been few phenomena wherein the quantum nature is
% apparent\cite{Fox}. 
% The interferometer we use is quite versatile and can be used to
% measure the fluctuations of various surfaces. In this work, we make
% direct measurements of the surface thermal fluctuation spectra of
% simple liquids. 
Interferometry is a most precise tool for detecting small
displacements and hence is used in a very broad range of areas in
physics, from microscopic spectroscopy such as ours, structural
measurements of optical elements to
astrophysics\cite{astroInterferometry} and arguably the most sensitive
measurement, the attempts to detect gravitational
waves\cite{gwRevs,LIGO2011}.  Surface thermal fluctuations have been
measured with interferometry previously, on mirrors using high power
lasers\cite{mirror} and on liquid drop surfaces attached to fiber
tips\cite{mitsuiFiber}. Spectral properties of surface waves on simple
liquids have also been measured by using them as gratings
\cite{rip1,ripplonExp}. However this approach is difficult to apply to
dissipative liquids such as oil, since they do not create well defined
waves that act effectively as gratings.  Spectra of surface
inclination fluctuations have been measured\cite{mitsui1,Tay,am1,am2}
using the surface as an optical lever.
% \cite{opticalLever0}. 
Also, exceptionally large surface fluctuations due to low surface
tension have been observed using other methods\cite{giant}.

In this work, we measure surface thermal fluctuation spectra of simple
liquids over a wide frequency range ($\rm 1\,kHz\sim40\,MHz$) and down
to few orders of magnitude below the shot noise level, using Michelson
interferometry. While these spectra reflect fundamental physics
principles of liquids and are interesting from a physics
perspective\cite{rip1,ripplon,ripplonExp}, the spectra we obtain have
not been done so previously, to our knowledge. The main reason for
this is that these fluctuations are small and are buried under the
shot noise.
%  In our case, the quantum noise reduction is of crucial
% practical value in this regard.  
% To reduce the noise in the
% measurements this far, we make essential use of the quantum nature of
% light. 
It is usually believed that the shot noise sets the limit for the
signal-to-noise ratio that can be obtained under normal
circumstances\cite{gwRevs,Fox,LIGO2011}, with the exception being
measurements involving sub-Poissonian photon statistics. We show both
theoretically and experimentally that the standard Michelson
interferometer, when combined with the quantum nature of light can
achieve signal-to-noise ratio unlimited by the shot noise.  In
essence, while both thermal fluctuations and shot noise are random,
the former are classical in their origin and the latter has
fundamentally quantum nature and their difference enables us to
separate these two.
% ; the quantum nature of light serves as a practical
% advantage in our method.
%
By being able to analyze the thermal fluctuation spectra to such
precision, their theoretical understanding can be examined in detail.
The traditional hydrodynamical description of simple liquids works
well for water, but some deviations from it for liquids such as oil
are observed at high frequencies.
% above 1\,MHz.

We now summarize briefly the general principle behind the noise
reduction employed here, which is not limited to optical measurements
nor to shot noise\cite{am1,am2}: In analyzing weak signals, if
the signal has a definite periodicity, we can accumulate data in
accordance with the period to suppress the noise.
% attain a high signal-to-noise ratio. 
However, in cases where there is no definite periodicity, it is
difficult to separate out random signals, such as thermal
fluctuations, from random noise. In fact, given a single measurement
$D_1=S+N_1$ of a random signal, there is no way, even in principle, to
distinguish the signal $S$ from the noise $N_1$, which inevitably
occurs. Shot noise is a typical example of such random noise. To
overcome this obstacle, we perform an additional measurement of the
same signal $D_2=S+N_2$ at the same time, whose noise $N_2$ is
independent of $N_1$. Then, by taking the correlation of $D_{1,2}$ and
averaging over time, we obtain
\begin{equation}
    \label{eq:corr}
    \vev{\overline{\tilde D_1}\tilde D_2}\longrightarrow \vev{|\tilde
      S|^2}
    \qquad
    ({\cal N}\rightarrow\infty)\quad,     
\end{equation}
% $\vev{\overline{\tilde D_1}\tilde D_2}\rightarrow \vev{|\tilde S|^2}\
% ({\cal N}\rightarrow\infty)$, 
where $\cal N$ is the number of
averagings and tildes denote Fourier transforms\cite{am1,am2}.  While
the above principle is simple, the crucial point is to arrange
multiple measurements of the same signal in such a way as to ensure
that the noise in them are uncorrelated.
%   The crucial point for this method to be
% applicable is that the noise in the measurements are independent.
% Here,
% we made use of the fact that since the measurements are independent,
% the correlation of the uncorrelated random noise averages to zero.

\begin{figure}[htbp]
    \centering
   \includegraphics[width=8cm,clip=true]{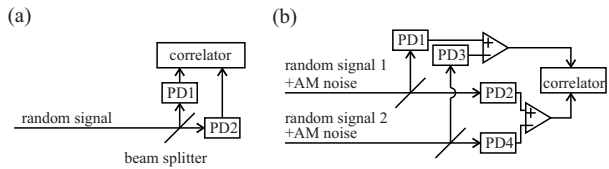} 
   \caption{(a)By taking the correlation of two photodetector (PD1,2)
     measurements, we separate the signal from the uncorrelated random
     noise. %
     (b)To eliminate AM noise that inevitably occur, we additionally
     use the differential detection of random signal 1 and 2, by adding
     PD3,4. Random signals 1,2 can either be phase inverted or
     uncorrelated.
% (b)To eliminate AM noise that inevitably occur, we additionally use
% differential detection, by adding PD3,4.
   }
    \label{fig:setup1}
\end{figure}

To measure surface fluctuations, laser light is shone on the surface
and the reflected  light is used as a signal for measurement. 
The quantum property of light plays a critical role in our experiment;
the beam splitter randomly partitions the photons from the single
signal beam to one of the two detection systems
(\figno{setup1}(a)). The photons giving rise to the shot noise is
random, when a coherent light source is used. Therefore, the noise in
the two detectors are uncorrelated and is eliminated when we compute
their correlation, Eq.~\eqnn{corr}. It is important to note that
splitting the beam by itself does not guarantee the independence of
the noise in the two measurements.  Had we used a squeezed light
source, for instance, the photons in the two measurements would have
been correlated, so that the noise could not have been eliminated by
using the correlation of detector measurements. Therefore, our
approach is in contrast to those that use squeezed light sources to
obtain sub-shot noise
measurements\cite{Caves81,squeezed,LIGO2011}. Using squeezed light
sources, a reduction of the shot noise by a factor of two have been
achieved and a light source squeezing factor close to 20 has been
attained\cite{squeezeSource}.

% To implement this principle in optical measurements, we make two
% measurements by splitting the light beam and using two photodetectors,
% as shown in \figno{setup1}(a). 
% While this is not sufficient by itself,
% this is a critical part in obtaining two measurements with
% uncorrelated noise, as explained below.
% Since these photodetector measurements
% are independent, uncorrelated random noise, including shot noise, can
% be eliminated. 
The signals obtained from the photodetectors (\figno{setup1}(a)), in
practice, additionally contain amplitude modulation (AM) noise from
the light source.  This is common to the two measurements and hence
will not be eliminated by taking their correlation. To reduce this
noise, we further employ differential detection (\figno{setup1}(b)).

\begin{figure}[htbp]
    \centering
   \includegraphics[width=8cm,clip=true]{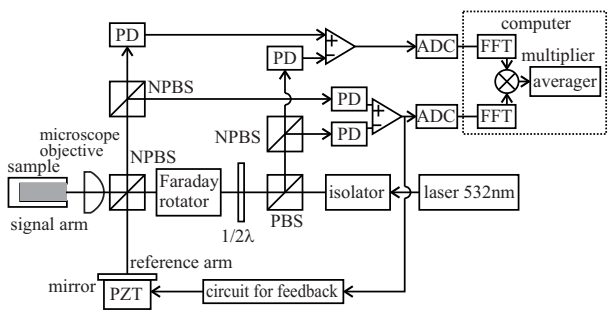}
   \caption{Experimental setup that incorporates noise reduction due
     to correlation and differential detection shown in
     \figno{setup1}. %
     NPBS:  nonpolarizing beam splitter.
     % PBS: polarizing beam splitter, , PZT: Piezo-electric transducer.  
   }
    \label{fig:setup2}
\end{figure}
Our Michelson interferometry setup that incorporates these principles
is shown in \figno{setup2}. The light source for this experiment is a
solid-state diode-pump laser (SAPPHIRE SF-532-50, Coherent) with a
wavelength of $\lambda$=532 nm arranged so that the power is 500
$\mu$W each at the sample and at the reference mirror.  A typical
measurement we show below requires a few seconds of measurement time.
% To reduce the frequency modulation (FM) noise,
% an external cavity was used for stabilizing the laser to the spectrum
% width of 40\,kHz\cite{laserStab}. 
To implement the differential detection, a Faraday rotator is used for
separating out the light reflected back from the sample at the
polarizing beam splitter (PBS). The lengths of the signal and
reference arms in the Michelson interferometer are adjusted to be
equal within 1\,$\mu $m, since a larger difference increases frequency
modulation noise effects. To obtain maximal sensitivity in detecting
sample surface displacements, the reference arm length is adjusted
dynamically by a piezo-electric transducer (PZT) with feedback.
In the Michelson interferometer, 
% care has to be taken to ensure that
% 
the light at the reference mirror and that at the sample need to have
the same properties
%  In our case, they are both planar in the focal plane of
% the respective objective lenses. 
or the destructive interference within the beam spot will reduce the
signal significantly.
Light reflected by the reference mirror and the sample is detected by
four photodiodes (S5973 Hamamatsu Photonics, Japan). The signal
currents obtained through differential detection are digitized by an
analog to digital converter (ADC, 8 bit, 125 Ms/s, PicoScope 5203,
Pico Technology). Their Fourier transforms and the correlation in
Eq.~\eqnn{corr} are calculated using a computer.

% details about calib?

%
Two light sources and two sets of detectors can also be used to obtain
two independent measurements and extract signals at sub-shot noise
levels from their correlation. This was the approach used for
inclination fluctuation spectra of surfaces\cite{am1,am2}.  In this
work, by using a single light source, we make minimal modifications to
the classic Michelson interferometer, essentially by just adding an
additional beam splitter at the photodetector, to attain sub-shot
noise measurements. This leads to a simple elegant setup. Using a
single light source is not only simpler, but has an important
practical advantage; to achieve maximal sensitivity in the
interferometer, the path lengths need to be adjusted according to the
wavelength of the light source, which is difficult to attain for
multiple wavelengths simultaneously.
% Furthermore, using a single light source leads to a simpler setup than
% using multiple sources since we do not need to deal with the
% complications of dealing with two wavelengths, such as including
% dichroic mirrors, taking care of dispersion or adjusting optimal beam
% path lengths for interference not just to one but to multiple
% % wavelengths.
% \comment{ denigrate 2 sources, complicated, two alignments, dichroic,
%   etc. not only that, in michelson, probably almost impossible to
%   adjust arm lengths to optimize sensitivity for multiple wavelengths.}

The spectral function of thermal surface fluctuations for a simple
liquid is determined from hydrodynamical considerations by its density
$\rho$, surface tension $\sigma$ and viscosity $\eta$ and is
\begin{equation}
    \label{eq:ripplon}
    P(k,\omega)={\kb T\over \pi}
    {ku^2\over \rho\omega^3}\Im\left[(1-iu)^2+y-\sqrt{1-2iu}\right]^{-1},
\end{equation}
% $ P(k,\omega)={\kb T } {ku^2/(\pi
%   \rho\omega^3)}\Im\left[(1-iu)^2+y-\sqrt{1-2iu}\right]^{-1} $%
where $u\equiv{\rho\omega/( 2\eta k^2)},y\equiv {\rho\sigma/(4\eta^2
  k)}$\cite{Bouchiat}. Here, $k,\omega$ are the wave number and the
angular frequency of the surface wave. Through Michelson
interferometry, we measure fluctuations perpendicular to the liquid
surface and its spectrum can be computed analogously to the
inclination fluctuation spectra\cite{am1,am2}  as
\begin{equation}
    \label{eq:dispH}
    \spec(f)=2\int_0^\infty dk\,k\,e^{-\bsize^2k^2/8}P(k,2\pi f)\quad,    
\end{equation}
%
% $S_{\rm  h}(f)=2\int_0^\infty dk\,k\,e^{-\bsize^2k^2/8}P(k,2\pi f)$,%
Here, $f=\omega/(2\pi)$ and $b$ is the beam diameter.
% and the beam size is $3\,\mu$m.
Surface waves with wavelengths larger than the sample size are cut off
so that the effects of gravity can be ignored. Also,
surface waves with wavelengths smaller than the beam spot size is
suppressed due to averaging.
% This is to be expected since properties of waves with
% wavelengths smaller than that of light can not be obtained from far
% field measurements.  
In the measurements below, the size of the surface sample is 2.2\,mm
in diameter and
% 1.37um-> 0.968um, 1.35um->0954um
$b=0.96\,\mu$m.

\begin{figure}[htbp]
    \centering
   \includegraphics[width=8cm,clip=true]{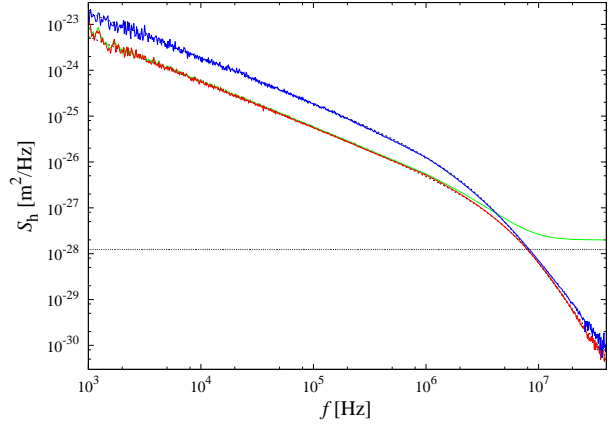} 
   \caption{(Color online) Experimentally observed surface height
     fluctuation spectra for water (red) and ethanol (blue). Water has
     smaller fluctuations. Respective theoretical spectra are also
     shown (black, dashed), which agree with the experimental results
     and are almost invisible.  For comparison, observed data for a
     single differential detection without using the correlation
     Eq.~\eqnn{corr} is shown for water (green) which is clearly
     dominated by the shot noise at higher frequencies.  The
     theoretical value for the shot noise level, Eq.~\eqnn{shotNoise},
     is also indicated (black, dotted). }
    \label{fig:waterEthanol}
\end{figure}
\begin{figure}[htbp]
    \centering
    \includegraphics[width=8cm,clip=true]{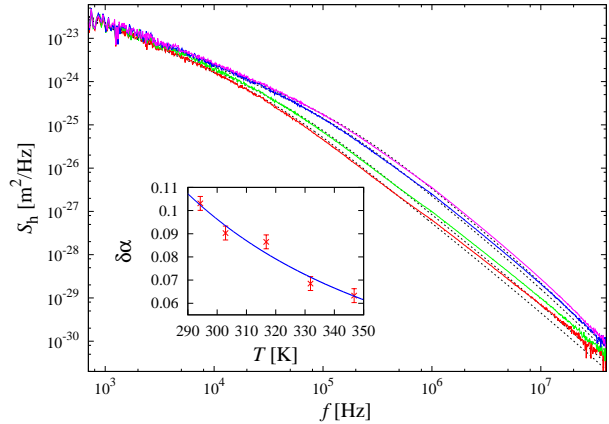} 
    \caption{(Color online) Experimentally observed surface
      fluctuation spectra for oil at various temperatures (red, green,
      blue, magenta from lower to higher temperatures). Fluctuations
      are larger at higher temperatures. Their respective theoretical
      spectra, Eq.~\eqnn{dispH}, are also shown (black,
      dashed). (Inset)Temperature dependence of the deviation  from
      theory of the
      fall off at high frequencies ($\gtrsim1\,$MHz).
%       ($\delta\alpha$).
    }
     \label{fig:oil}
\end{figure}
We compare the measured spectra against their theoretical predictions,
Eq.~\eqnn{dispH}, for water and ethanol in \figno{waterEthanol}.  The
physical properties of water, ethanol are well known for a given
temperature and are $(\rho\,{\rm [kg/m^3]}, \sigma\,{\rm
  [kg/s^{2}]},\eta\,{\rm [kg/( m\cdot s)]},T[{\rm
  K}])=(1.0\times10^3,7.5\times10^{-2},1.7\times10^{-3},275)$,
$(0.80\times10^3,2.3\times10^{-2},1.7\times10^{-3},278)$,
respectively. Given the temperature, these properties and the spectral
function $P(k,\omega)$ are completely determined. The integrated
spectrum $S_{\rm h}(f)$, Eq.~\eqnn{dispH}, is uniquely determined by
those properties and the beam size.
Water and ethanol are cooled to suppress evaporation which can cause
the sample surface to leave the focal plane.
%  and greatly reduce the
% signal-to-noise ratio.
%
For water surface fluctuations, the theoretical formula derived from
hydrodynamic considerations agree essentially perfectly with the
observed spectrum.  These measurements confirm the theoretical
spectrum over a wide frequency range. For ethanol, good agreement is
seen, except for the slightly slower fall off at frequencies above
10\,MHz. We come back to this issue for the case of oil surface
fluctuations.

Oil (Olympus immersion oil AX9602) surface thermal fluctuation spectra
were obtained at various temperatures and compared to the theoretical
spectra Eq.~\eqnn{dispH} in \figno{oil}.  There is a qualitative
difference from the spectra of water and ethanol surface fluctuation
spectra in \figno{waterEthanol}, due to the high viscosity of the
liquid. In particular, the spectral density decays much more slowly
than that of water and oil at higher frequencies. For the oil, the
temperature dependence of $\sigma,\eta$ are not known so that they had
to be deduced from the spectra and were found to be $(\rho\,{\rm
  [kg/m^3]}, \sigma\,{\rm [kg/s^{2}]},\eta\,{\rm [kg/( m\cdot
  s)]},T[{\rm K}])=(0.92\times10^3,3.2\times10^{-2},0.15,294)$,
$(0.91\times10^3,3.1\times10^{-2},0.096,303)$,
$(0.89\times10^3,2.9\times10^{-2},0.037,332)$,
$(0.88\times10^3,2.7\times10^{-2},0.025,347)$. %
$\sigma$, $\eta$ are determined to ten percent from the spectra.
For oil, we see that the fluctuation spectrum changes with the
temperature, as it should. Most of this change is caused by the
decrease in the viscosity with higher temperatures, making the
fluctuations larger. 

While the fluctuation spectra of liquids we study are usually regarded
as well understood, the observed higher frequency ($\gtrsim1\,$MHz)
fall offs in the spectra for oil and ethanol are slightly slower than
their theoretical predictions, especially at lower temperatures. This
is possibly due to the more complex nature of the liquids that can not
be explained just from the hydrodynamical considerations of simple
liquids.
The slower fall off of the spectrum has also been observed for solid
materials and a gradual transition to such a dependence was seen for
complex fluids\cite{am1}.  We obtain the dependence $S_{\rm h}(f)\sim
f^\alpha$ for higher frequencies ($10^6-10^7$\,Hz) and its deviation
$\delta \alpha$ from the theoretical value in Eq.~\eqnn{dispH}. In
\figno{oil}(inset), the temperature dependence of $\delta\alpha$ is
shown. The dependence can be reasonably well described by
$\delta\alpha=C\exp(U/k_{\scriptscriptstyle\rm B}T)$ with
$U=7\,$kJ/mol. This energy scale $U$ is comparable to 
% the binding
% energy of oil molecules, as evidenced in 
the latent heat for oil, which is consistent with more complex
behavior, such as molecular interactions causing visible effects in
the spectrum.
%
% when compared to simpler liquids such as water 
% It seems difficult to obtain the spectral information of highly
% dissipative liquids in this regime from other approaches, since the
% dissipative nature prevents the formation of well defined gratings
% from the surface waves.
% and the spectral measurements are
% well below the shot noise level.

In \figno{waterEthanol}, we also included the results from one
differential measurement of water surface fluctuations, in which the
shot noise level is clearly visible.  Theoretically, the shot noise
level in the spectrum
is
% consistent with its theoretical value, which 
% can be derived as
\begin{equation}
    \label{eq:shotNoise}
    e{\lambda^2\over 32\pi^2}{1+r^2\over rI_{\rm PD}}\Delta f
    \quad, 
\end{equation}
% $e\lambda^2/(32\pi^2)\cdot(1+r^2)/(rI_{\rm PD})\cdot \Delta f$, 
when the sensitivity in the interferometer is maximal.  Here, $r$ is
the ratio of the reflectivities of the sample surface and the
reference mirror, $I_{\rm PD}$ the signal photocurrent and $e$ the
electron charge.  As can be seen in \figno{waterEthanol} and \ref{fig:oil},
shot noise is clearly eliminated when the correlation of the two
differential measurements are taken. 
% Using the correlation,
% Eq.~\eqnn{corr}, we were able to make measurements down to three
% orders of magnitude below the shot noise level.
% 
The observed shot noise level in \figno{waterEthanol} is roughly twice
its theoretical value, Eq.~\eqnn{shotNoise}. This indicates that the
interferometer sensitivity is not at its theoretical maximum, whose
most likely cause is the aberration of the objective lens.
% the liquid surface is not precisely in the focal
% plane, which can give rise to some non-planarity in the waves at the
%  surface, 
%  leading to some unwanted destructive interference.

Let us briefly describe the physical properties of the spectra
Eq.~\eqnn{dispH}: When $16\sqrt2\pi\eta^3f/(\rho\sigma^2)\gtrsim1$,
the liquid can be regarded as being highly viscous so that any liquid
is dissipative at high enough frequencies. In the frequency region we
study, water, ethanol have low and oil has high viscosity. The spectra
have qualitatively different $f$ dependence for low and high
viscosities, as explained below\cite{am1,am2}. In both cases,
$\spec(f)\sim \kb T/(\sigma f)$ at low frequencies.  For a liquid with
low viscosity, the spectrum crosses over at $f\sim\sqrt{\sigma/(\rho
  b^3)}$ to $\spec(f)\sim \kb T\eta/(\rho^2b^5f^4)$. For a liquid with
high viscosity, the spectrum crosses over at $f\sim\sigma/(\eta b)$ to
$\spec(f)\sim \kb T/(\eta b f^2)$. While outside the region of our
measurements, at even higher frequencies, $f\gtrsim \eta/(\rho b^2)$,
the spectrum changes to $\spec(f)\sim \kb T\eta/(\rho^2b^5f^4)$. For
both low and high viscosities, the behavior of $\spec(f)$ is dominated
by $\sigma$ at low frequencies and is independent of $\eta$ since the
dynamical time scale is relatively large. At high frequencies, the
behavior is governed by $\eta$ and is independent of $\sigma$, to
leading order.  These crossovers seen in the spectra are similar to
those in the surface inclination fluctuation spectra and arise from
the properties of $P(k,2\pi f)$,
Eq.~\eqnn{ripplon}\cite{am1,am2}. However, the spectra differ
qualitatively. In particular, $\spec(f)$ has a distinctive $1/f$
dependence at lower frequencies and is independent of the beam
diameter $b$ to leading order. Consequently, the height fluctuation
measurements are more sensitive than the inclination fluctuation
measurements at lower frequencies. This is natural since the
inclination fluctuations due to longer wavelength fluctuations are
smaller for a fixed beam size.  At high frequencies, the cutoff
frequency determined by $b$ becomes important, since it is responsible
for suppressing the shorter wavelength modes.

In this work, we have proposed and implemented the Michelson
interferometer with noise reduction at sub-shot noise levels.
We make multiple measurements of a single signal light. The shot noise
in them are independent due to their quantum nature, hence the noise
can be eliminated through their correlation and the signal below this
noise level can be extracted.
While both random, the distinct difference between the thermal
fluctuations and the shot noise, which are classical and quantum in
their origins, allow for their separation. It would be interesting to
understand the behavior when the method is applied to quantum
fluctuation measurements with shot noise.
At present, some applications such as quantum cryptography and quantum
random number generation exist, yet there have not been many cases
where the quantum property of light has been of practical
value\cite{Fox} and it is satisfying to find it provide a practical
advantage here.
% ; the quantum nature of light 
%
The measurement requires only a small light power, a relatively short
time and a small sample surface ($500\,\mu$W, few seconds, diameter
1\,$\mu$m, in this work). Our noise reduction can also be added to
higher power measurements such as those in \cite{mirror} to reduce the
noise further.  Longer measurement times can lead to higher
resolution, more averaging and less statistical error.  Interferometry
is perhaps the most commonly used method for precision measurements
and shot noise can often be the major limiting factor in its
accuracy\cite{Fox,LIGO2011,gwRevs}.
% Our approach can be used as long
% as the signal can be averaged. 
We believe that our method can be applied in various situations in
which precision, low power or short measurement time is required.

Let us discuss the limitations of our approach: The approach is
effective when the number of averagings ${\cal N}=\Delta f T$ is
large, where $\Delta f$ is frequency resolution in the spectrum and
$T$ is the total measurement time. The relative error in the spectrum
is $1/\sqrt{\cal N}$. $\Delta f$ needs not be the same across the
spectrum and if a same relative resolution is used, this statistical
error is smaller at higher frequencies.  While this integration is
effective for stationary signals, the situation is more subtle for
transient signals such as gravitational waves radiated from a single
event. In such situations, the restriction that the measurement time
needs to be within the time when the signal is present can be
demanding.  The duration of the signal needs to be long enough
compared to the inverse of the desired frequency resolution, to reduce
the noise using the correlation of measurements. 

\end{document}